\documentclass[conference]{IEEEtran}
\IEEEoverridecommandlockouts
\usepackage{cite}
\usepackage{amsmath,amssymb,amsfonts}
\usepackage{algorithmic}
\usepackage{graphicx}
\usepackage{textcomp}
\def\BibTeX{{\rm B\kern-.05em{\sc i\kern-.025em b}\kern-.08em
    T\kern-.1667em\lower.7ex\hbox{E}\kern-.125emX}}

\usepackage[table,xcdraw]{xcolor} 
\usepackage{listings} 
\usepackage{caption} 
\usepackage{subcaption} 
\begin{document}

\title{Large Language Models as Visualization Agents for Immersive Binary Reverse Engineering\\
}

\author{\IEEEauthorblockN{1\textsuperscript{st} Dennis G. Brown}
\IEEEauthorblockA{\textit{Comp. Sci. and Software Eng.} \\
\textit{Auburn University}\\
Auburn, AL, USA \\
dgb0028@auburn.edu}
\and
\IEEEauthorblockN{2\textsuperscript{nd} Samuel Mulder}
\IEEEauthorblockA{\textit{Comp. Sci. and Software Eng.} \\
\textit{Auburn University}\\
Auburn, AL, USA \\
szm0211@auburn.edu}
}

\maketitle

\begin{abstract}

Immersive virtual reality (VR) offers affordances that may reduce cognitive complexity in binary reverse engineering (RE), enabling embodied and external cognition to augment the RE process through enhancing memory, hypothesis testing, and visual organization. In prior work, we applied a cognitive systems engineering approach to identify an initial set of affordances and implemented a VR environment to support RE through spatial persistence and interactivity. In this work, we extend that platform with an integrated large language model (LLM) agent capable of querying binary analysis tools, answering technical questions, and dynamically generating immersive 3D visualizations in alignment with analyst tasks. We describe the system architecture and our evaluation process and results. Our pilot study shows that while LLMs can generate meaningful 3D call graphs (for small programs) that align with design principles, output quality varies widely. This work raises open questions about the potential for LLMs to function as visualization agents, constructing 3D representations that reflect cognitive design principles without explicit training.


\end{abstract}


\section{Introduction}
\label{section:introduction}

Binary reverse engineering (RE) is the practice of understanding the meaning and capabilities of a binary executable from several sources of low-level evidence. It is a cognitively-demanding sensemaking task, requiring analysts to build and test hypotheses about program properties while synthesizing higher-level structures from incomplete and uncertain information. The process often relies heavily on intuition, experience, and tool-assisted manual exploration.

Building on prior work \cite{Brown2024} that identified cognitive principles relevant to reverse engineering (e.g., embodied cognition, external memory, and cognitive load management) and implemented them in our Cognitive Binary Reverse Engineering (CogBRE) VR tool, this study investigates whether large language models (LLMs) can act as visualization agents. Specifically, we examine their ability to generate and organize 3D representations of program artifacts in ways that align with human perceptual and cognitive design principles, hypothesizing that LLMs possess sufficient latent knowledge to produce visualizations that support sensemaking without explicit training.

In this paper, we present early-stage research in which an LLM can invoke reverse engineering tools and process their output to generate spatial visualizations in the immersive workspace that are shaped by spatial and perceptual cues that align with task goals. This work is best understood as an early-stage design study with elements of formative evaluation. The contribution of this work includes the design and implementation of an LLM-integrated immersive RE system and a pilot evaluation testing the hypothesis that LLMs show potential to act as cognitive-aware visualization agents without explicit training.

\begin{figure}[tbp]
    \centerline{\includegraphics[width=0.49\textwidth]{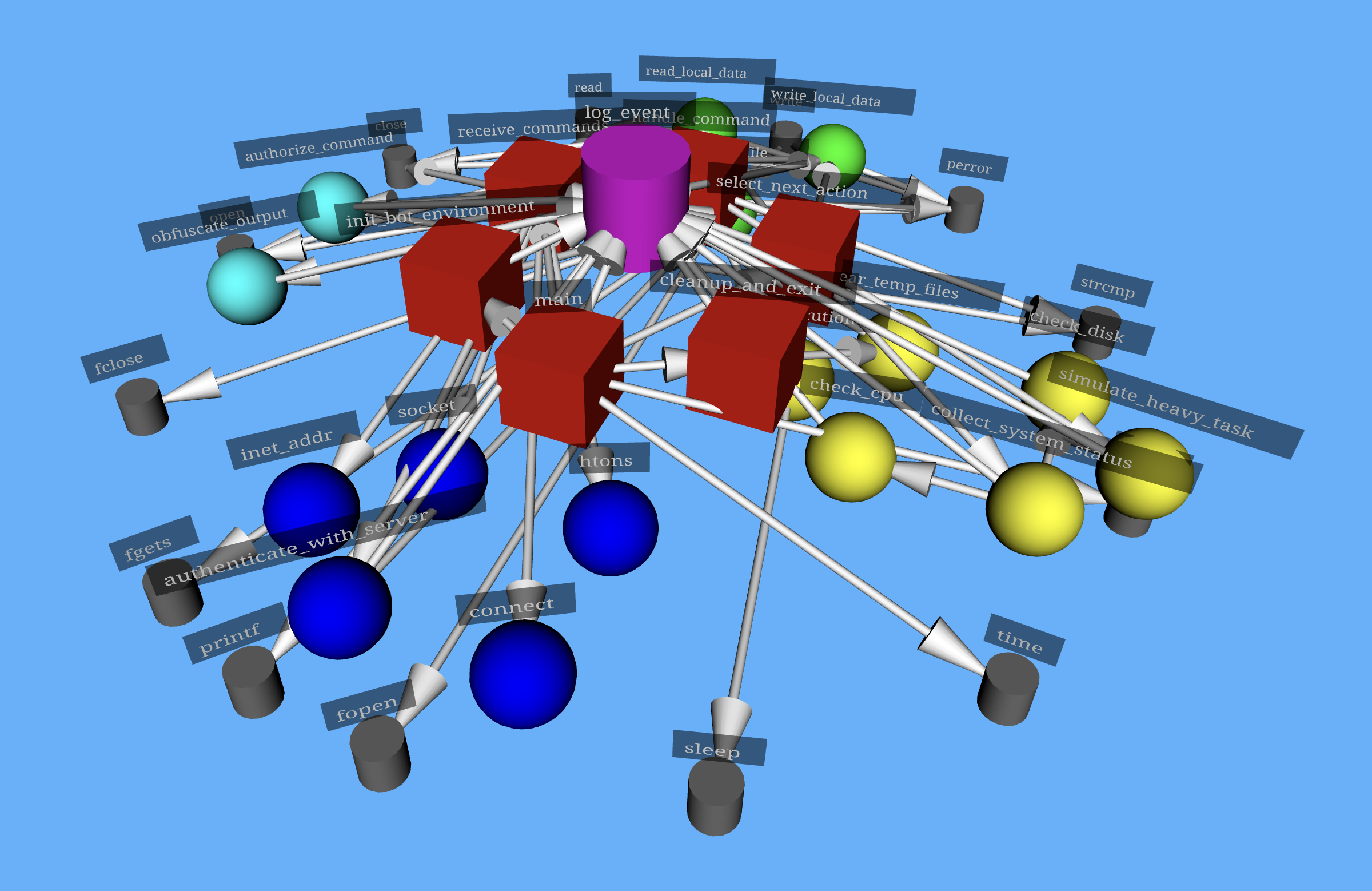}}
    \caption{Call graph designed by an LLM after performing self-directed discovery of a binary program simulating a remote-controlled bot. Note overall layout in hemispherical shape with higher-level functions in the center radiating out to groups of related functions and system calls on the periphery.}
    \label{fig:teaser}
\end{figure}

\section{Related Work and Motivation}
\label{sec:background}

Prior work in cognitive psychology and perceptual design provides principles for effective visualizations. Given the extensive inclusion of technical, scientific, and design documentation in typical LLM training corpora, it is reasonable to expect that patterns reflecting cognitive and perceptual design principles are embedded in LLM behavior. We do not claim that LLMs reason about perception or cognition in a principled manner; rather, we hypothesize that they can heuristically generate visualizations that reflect effective design patterns. Effective visualizations should align with user tasks~\cite{Munzner2014} and should also exploit pre-attentive processing (color, shape, position), use spatial organization to support memory and reasoning, and apply grouping to reduce cognitive burden~\cite{Ware2012}. Visualizations should maximize information and minimize noise, avoid unnecessary complexity, and be quick to comprehend~\cite{Tufte1983}. 

While current binary RE tools (e.g., Ghidra~\cite{Ghidra2023}) offer visualizations, they primarily provide only 2D graphs in conventional layouts that present a limited selection of qualities of the binary file per graph. Prior work in immersive analytics and software visualization has demonstrated the potential of spatial layouts, persistent visualizations, and embodied interaction to support sensemaking tasks~\cite{Lisle2021, Batch2020, Fittkau2015, Romano2019}. 
Other recent work explores LLMs for information visualization, such as using LLM agents to generate charts for varied contexts and tasks~\cite{Cui2025} and determining and modeling the design preferences expressed by LLMs~\cite{Wang2025}.

However, we have found no prior work that uses LLMs to autonomously create or organize 3D visual structures aligned with human cognitive affordances for the task of sensemaking in binary RE. LLMs may possess latent capabilities for generating useful, cognitively-aligned visualizations. Such visualizations could assist in reducing extraneous cognitive load, managing workspace organization, and supporting goal iteration within a 3D space that leverages embodied and external cognition.

\section{System Overview: LLM-Augmented CogBRE}
\label{sec:implementation}

\begin{figure}[tbp]
    \centerline{\includegraphics[width=0.5\textwidth]{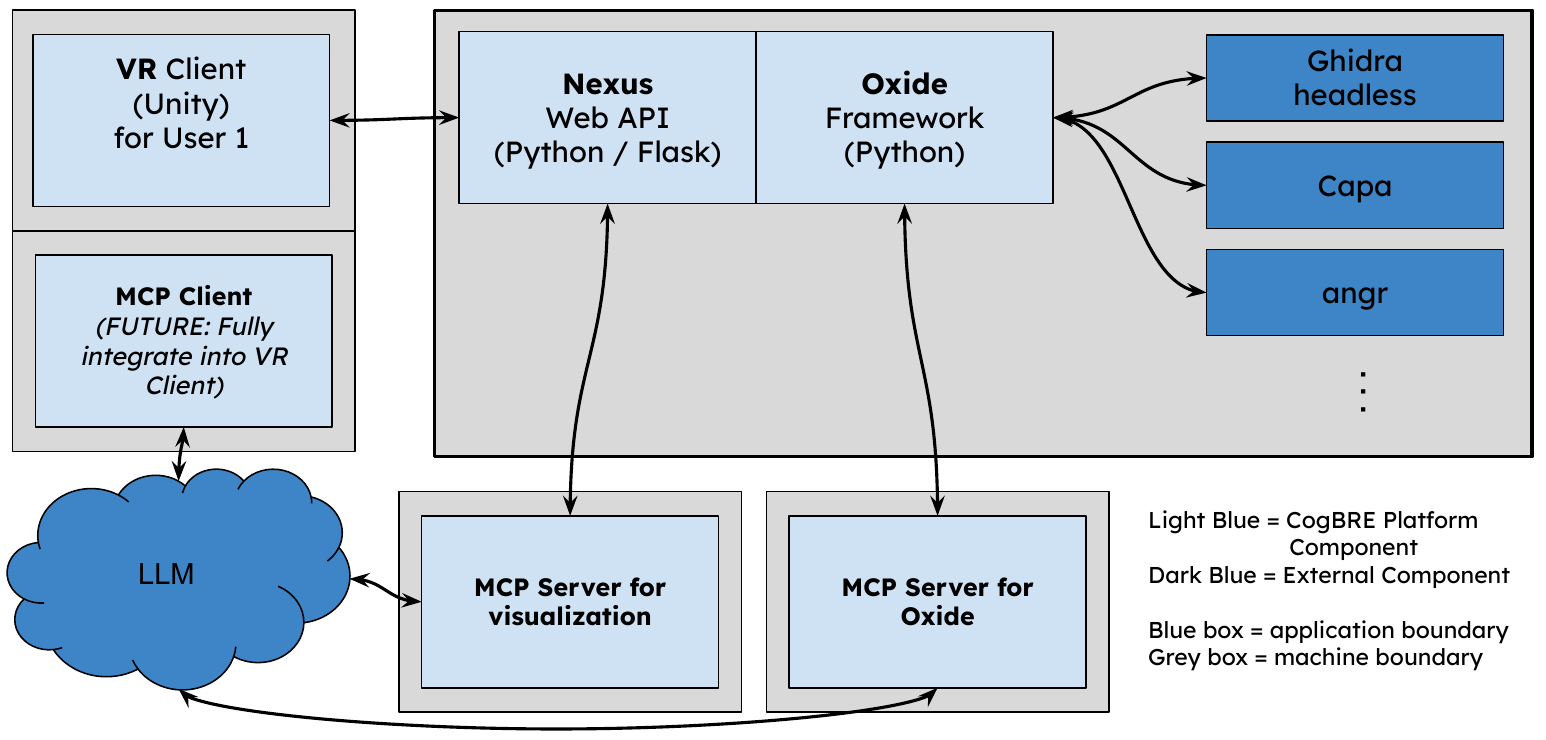}}
    \caption{CogBRE System Architecture with LLM and MCP.}
    \label{fig:architecture}
\end{figure}

To test our hypothesis, we updated CogBRE by incorporating an intelligent agent driven by an LLM. At the core of this integration is the Model-Context-Protocol (MCP) framework~\cite{MCP2025}, which structures interactions between the LLM, tools, and immersive environment. Figure ~\ref{fig:architecture} shows the updated architecture. This architecture enables the LLM to function as an informed collaborator, capable of tool use and immersive world manipulation. 

The system consists of six key components:
(1) The CogBRE VR client, where the user interacts in immersive space with a tracked head-mounted display and hand controllers; (2) Nexus, a broker between the VR client and data providers; (3) Oxide, a modular RE analysis engine~\cite{mulder2014}; (4) The MCP visualization server, which translates LLM directives into spatial commands for the VR client that are brokered by Nexus; (5) The MCP Oxide server, which provides tools to answer LLM questions about binary files through calls to the Oxide API (and indirectly, a suite of common RE tools); and (6) An MCP client, which mediates between the LLM and the rest of the system. For this work, we employed OpenAI's GPT-4.1 and o4-mini~\cite{OpenAI2025} LLMs. When responding to a prompt, the LLM provides JSON-based scene geometry to the MCP visualization server.

\section{Evaluation Design}
\label{sec:eval_design}

In order to gain insight into our hypothesis that LLMs contain enough embedded knowledge to act as perceptual designers, we ran a pilot evaluation to rate visualizations generated by LLMs under varying conditions including level of guidance provided by the prompt. We employed subjective measures rated by humans (four graduate students studying binary reverse engineering) and objective measures calculated automatically. 

We set limits on the pilot evaluation to reduce confounding variables, learn enough to inform future work, and respect the time of our raters. Traditional 2D call graphs from standard RE tools served as a reference for correctness (``ground truth''). We sought to explore whether LLMs, under varying conditions, could produce layouts that plausibly align with cognitive and perceptual design principles and not to measure absolute performance improvement over traditional tools.

Each of the four raters received their own evaluation package with the LLM-generated visualizations (as static 3D models) and the LLM’s stated design reasoning. Packages included all 40 visualizations in random order with the guidance condition and LLM hidden. Raters were told the program being visualized and were given its source code and a traditional 2D call graph as ground truth.
While the system does produce models in the immersive VR environment, raters instead viewed the models in a desktop 3D model viewer to accommodate participants' availability. 

The subjective rating dimensions were inspired by Munzner, Ware, and Tufte as referenced in Section~\ref{sec:background} plus a measure of correctness against ground truth. Raters provided a Likert scale rating of 1 (worst) to 5 (best) in each of these dimensions: \emph{Clarity:} How easily can you understand what this visualization is showing? \emph{Task Fit:} Would this help you answer questions about the program? \emph{Spatial organization:} Are related elements grouped and arranged meaningfully in space? \emph{Cognitive load:} What is the mental effort required to interpret this visualization? (1 = most effort; 5 = least) \emph{Visual encodings:} Are size, shape, color, and position used effectively? \emph{Correctness:} Does it lack obvious errors or misleading information? 

Our objective metrics were chosen to quantify structural and perceptual properties of the LLM-generated call graphs. \emph{Edge crossings:} sum of crossings detected when projecting the graph to the XY and XZ planes, which reflects layout clarity and directly impacts the visual complexity of the graph. \emph{Spatial dispersion:} the average pairwise distance between nodes, which serves as a proxy for grouping quality: tighter clusters may indicate more interpretable layouts. \emph{Hierarchy depth:} reflects the structural complexity of the call graph and may influence how effectively the LLM captures the program's control flow. \emph{Color and shape diversity:} assess the variety of visual encodings applied to distinguish function roles or behaviors. \emph{Average and standard deviation of edge lengths:} relate to the uniformity and spatial balance of the layout. While these metrics align with visualization design principles known to influence cognitive load, task efficiency, and visual clarity, we will use them primarily to measure \emph{variation} between visualizations, not \emph{quality} of the visualization. 

\section{Prompting Methodology}
\label{sec:prompting_methodology}

A key question in building an LLM-driven visualization system is whether to constrain the model to known visual conventions or to allow it to freely explore the space of possible visual encodings. Fine-tuning an LLM on human-designed visualization patterns may produce more cognitively-grounded outputs (layouts that align with spatial reasoning principles and visualization best practices). However, such fine-tuning risks narrowing the model's generative capacity, potentially suppressing novel or serendipitous structures that may emerge from its broader training. Instead, we advocate for a middle ground: allow the LLM to remain flexible but guide it with prompt engineering. This design allows the model to explore novel spatial structures while remaining aligned with principles of human perception and task relevance. 

We designed a prompt for each of the 8 configurations in two parts.
Each starts with the language as shown in Listing~\ref{listing:promptintro}, giving consistent guidance to the LLM on how to query information for the call graph and directing which binary program to examine. We designed the prompt in an ad-hoc manner, iterating it until we saw no obvious improvements in results. 

\begin{lstlisting}[float,floatplacement=tbp,language={},label=listing:promptintro, caption={Common Prompt Introduction}]]
You are an assistant in an immersive virtual reality 
system designed to help users reverse engineer 
binary programs.
 
The user operates in 3D space and can view visual 
artifacts such as call graphs, control flow graphs, 
and text windows (slates) with function listings. 

Your overall goal is to use the available 
capabilities to help the user understand how the 
binary works.

You should use the tools provided to understand what 
functions are in the binary program, the 
capabilities of the functions, and review the 
decompiled pseudo-source code to understand what key 
functions do. You may need to make many tool calls. 
 
The binary file has ID = <insert ID here>

After you have your best understanding of the 
program's structure and purpose, build a 3D function 
call graph that is designed and organized in a way 
to convey the most meaning to the user. 
\end{lstlisting}

The \emph{guidance level} portion of each configuration has two possibilities, ``low'' and ``high,'' referring to the level of design guidance given in the LLM prompt to generate the call graph. The low guidance condition tests how the LLM behaves when only minimally framed with no cognitive design principles or layout heuristics: no mention of grouping, spatial layout, or design principles. This condition leverages only the latent design knowledge in the LLM. The remainder of the prompt for this condition is simply ``Explain your reasoning.'' The high guidance condition provides principles (not rules)  that represent plausible heuristics the LLM may have seen during training. The remainder of the prompt for this condition is shown in Listing~\ref{listing:highconclusion}.

\begin{lstlisting}[float,floatplacement=tbp,language={},label=listing:highconclusion, caption={High Guidance Prompt Conclusion}]]
To support the user's reasoning, try to:
- Group related elements spatially
- Use color or shape to distinguish different function types or behaviors
- Avoid unnecessary clutter or overlap
- Place important elements where they are easy to notice
- Label elements when that helps clarity
Explain your reasoning.
\end{lstlisting}

Our high guidance condition is grounded in widely-accepted principles from visualization and cognitive science. We draw on work referenced in Section ~\ref{sec:background}: Munzner’s nested design model~\cite{Munzner2014} to align visualizations with user tasks; Ware’s perceptual theory~\cite{Ware2012}, based on classical Gestalt principles, to account for preattentive processing, grouping, and cognitive load; and Tufte’s principles~\cite{Tufte1983} to emphasize clarity, density, and visual economy. Together, these sources provide a well-rounded framework to assess how varying levels of design guidance affect the quality and cognitive plausibility of visualizations generated by an LLM in immersive reverse engineering tasks. While this framework does not exhaust all possible dimensions of visualization design, it provides a practical and theoretically grounded basis for evaluating structural layout and cognitive alignment in the context of immersive static analysis.

Another configuration element is the binary program being examined. Raters examined two programs with differing characteristics to provide some generalization to our results. The first is ``hexdump'' from the Canonical Ubuntu~\cite{Canonical2023} 20.04 x86-64 distribution, compiled with optimization and therefore obfuscated function names. The second is ``v11,'' a custom-built simulation of a potentially malicious bot written in C. This program attempts no obfuscation within the source and was compiled without optimization for x86-64, which carried over original function names to the binary file. The first program is a typical program as found ``in the wild,'' while the second was specifically designed to exercise our system and to facilitate understanding by the raters. Both programs are small, with tens of functions, as we found that larger programs would hit the LLMs' context window and tokens-per-minute limitations. 

The final configuration element is the LLM (GPT-4.1 or o4-mini). This variation gauges the impact of LLM selection on call graph creation. We chose a flagship chat model and a cost-optimized reasoning model to represent common types of commercial LLMs; we limited the selection to two models to keep this pilot evaluation manageable and cost-effective. The three parts in our configurations (guidance level, model, and program) each have two options, yielding the 8 distinct configurations we mentioned earlier.

\section{Evaluation Results and Analysis}
\label{sec:results}

Using the previously-defined subjective and objective metrics, we evaluated the 8 configurations of LLM-generated call graphs (40 total). Table~\ref{table:results_summary} shows the mean ratings and coefficients of variation (CV) for each of the 8 configurations. Subjectively, graphs for ``v11'' were rated higher than those for ``hexdump,'' likely because the ``v11'' graphs always had meaningful function names carried over from the unoptimized binary, while ``hexdump'' function names were inconsistently inferred by the LLM; this finding may demonstrate the importance of semantic context in generating visualizations. Guidance level showed only a modest effect on average ratings (High = 3.00 vs. Low = 2.86), suggesting that LLMs benefit somewhat from informal perceptual design cues, but results were not conclusive. GPT-4.1 and o4-mini performed similarly on average. 

Performing t-tests on the objective measures, we found that high guidance significantly increased shape diversity (p = 0.01), average edge length (p = 0.02), and edge length variability (p = 0.05). These metrics suggest more visually expressive and differentiated layouts under high guidance, though not always better-organized. GPT-4.1 visualizations had significantly more color diversity than those from o4-mini (p $<$ 0.01), implying greater use of pre-attentive color channels, though this did not necessarily correspond to better ratings.

\begin{figure*}
    \centering
    \begin{subfigure}[b]{0.49\textwidth}
        \centerline{\includegraphics[width=\linewidth]{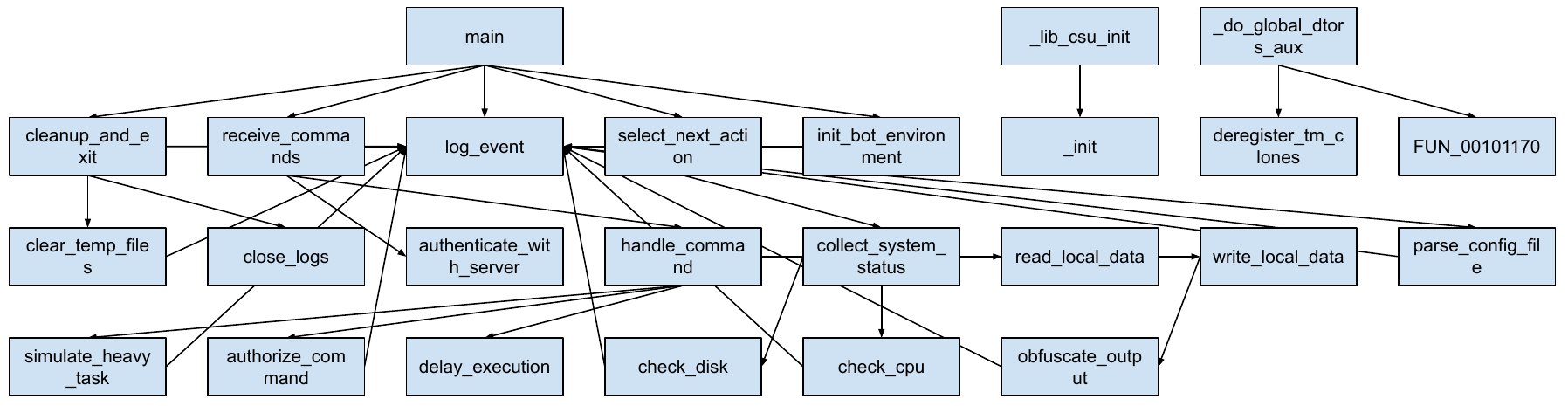}}
        \caption{Basic 2D call graph for v11 (without external calls)}
        \label{fig:v11_2d}
    \end{subfigure}
    \hfill
    \begin{subfigure}[b]{0.49\textwidth}
        \centerline{\includegraphics[width=\linewidth]{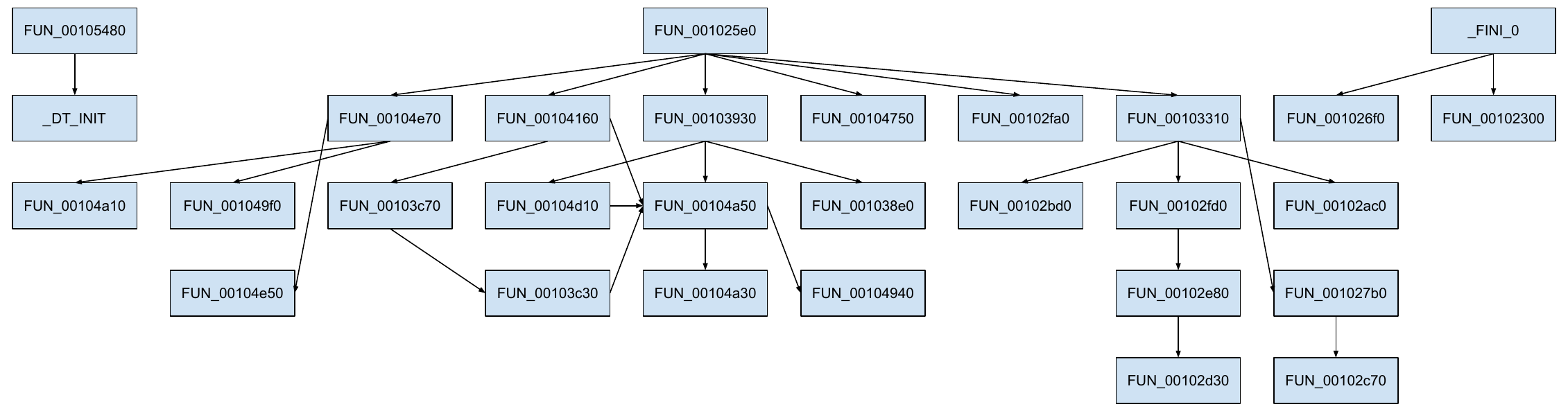}}
        \caption{Basic 2D call graph for hexdump (without external calls)}
        \label{fig:hexdump_2d}
    \end{subfigure}

    \vskip\baselineskip 

    \begin{subfigure}[b]{0.49\textwidth}
        \centerline{\includegraphics[width=0.75\linewidth]{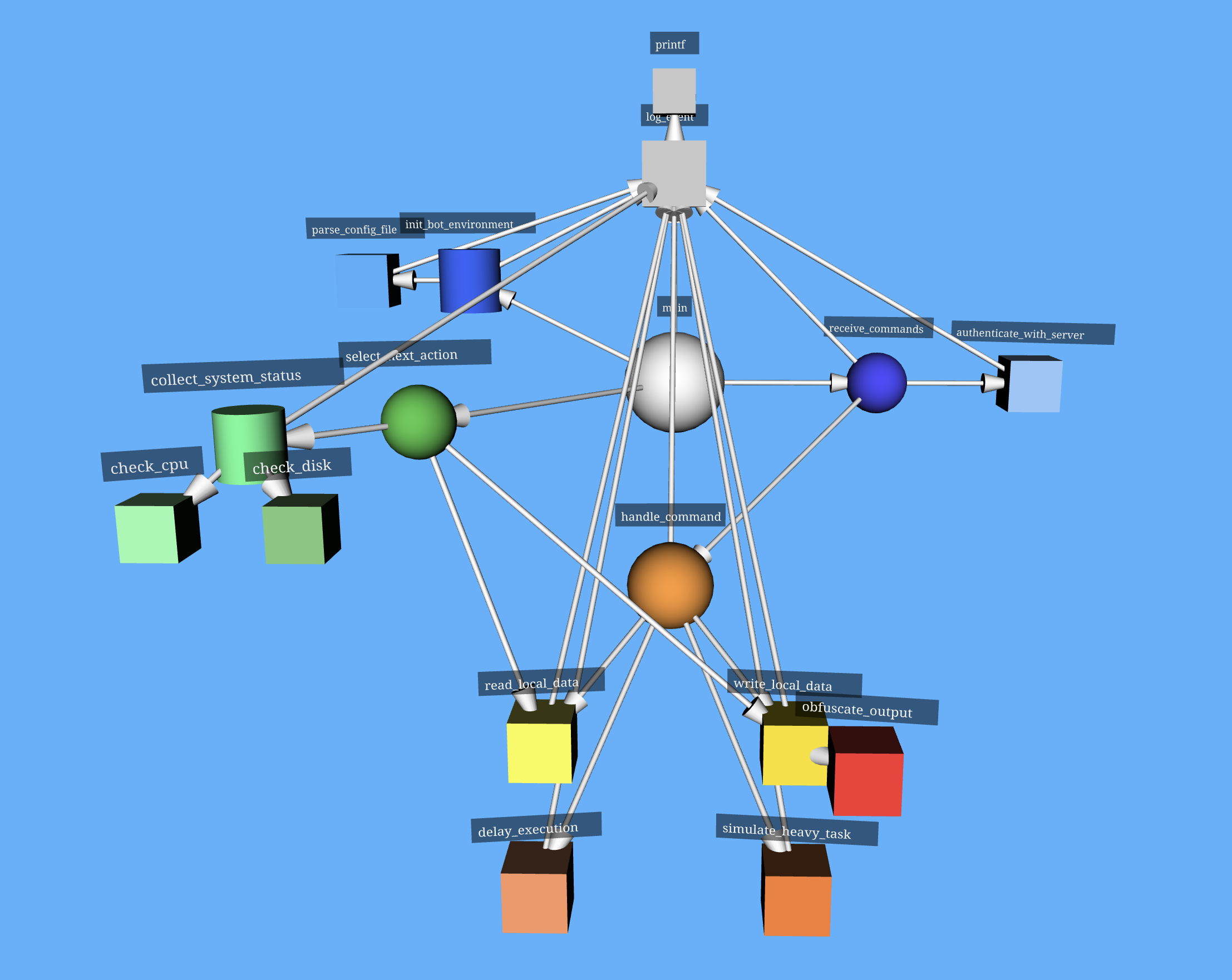}}
        \caption{Highest-rated call graph for v11 (subjective): High guidance, GPT-4.1}
        \label{fig:highest_sub_v11}
    \end{subfigure}
    \hfill
    \begin{subfigure}[b]{0.49\textwidth}
        \centerline{\includegraphics[width=0.75\linewidth]{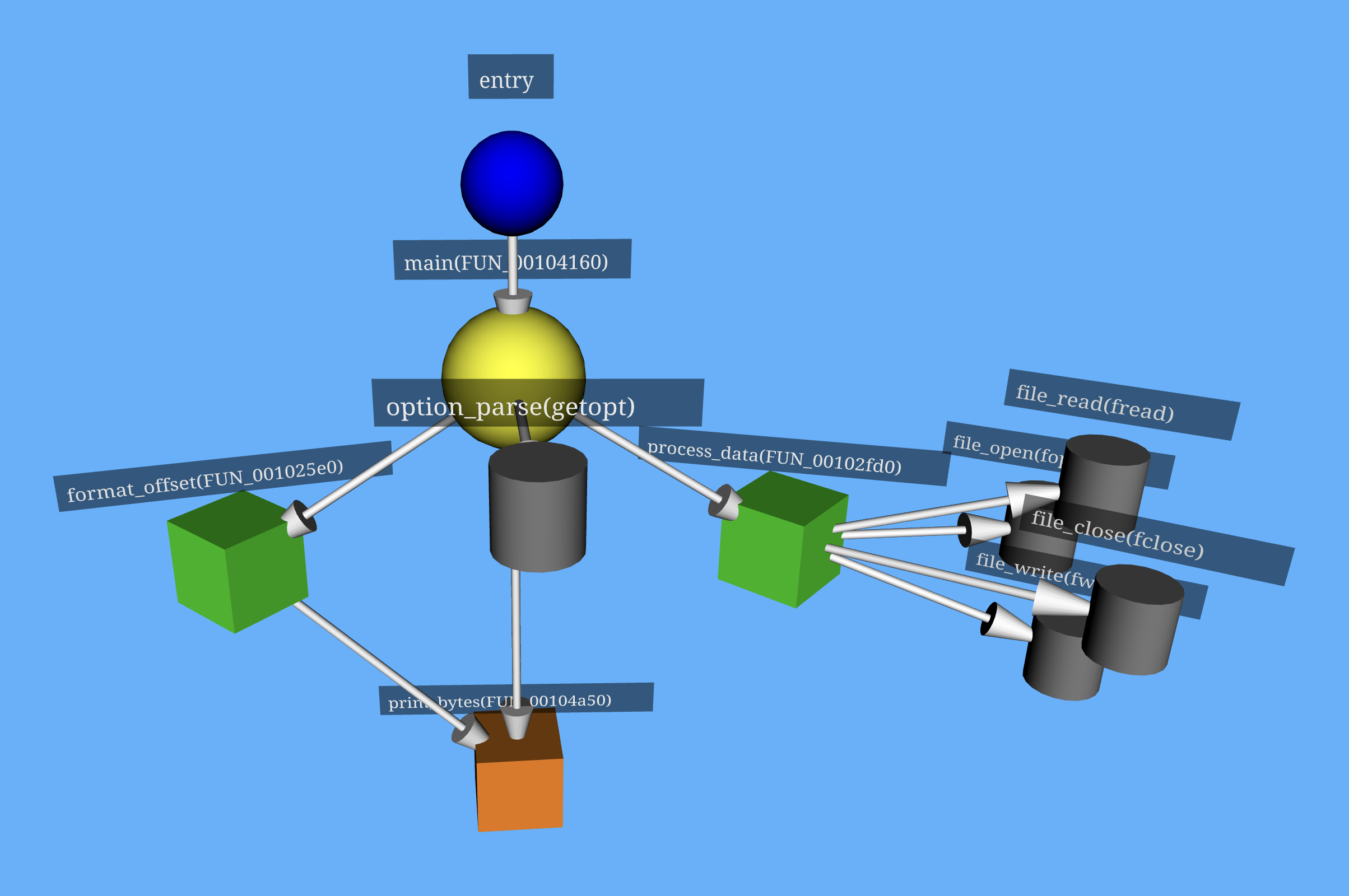}}
        \caption{Highest-rated call graph for hexdump (subjective): High guidance, o4-mini}
        \label{fig:highest_sub_hexdump}
    \end{subfigure}

    \vskip\baselineskip 

    \begin{subfigure}[b]{0.49\textwidth}
        \centerline{\includegraphics[width=0.70\linewidth]{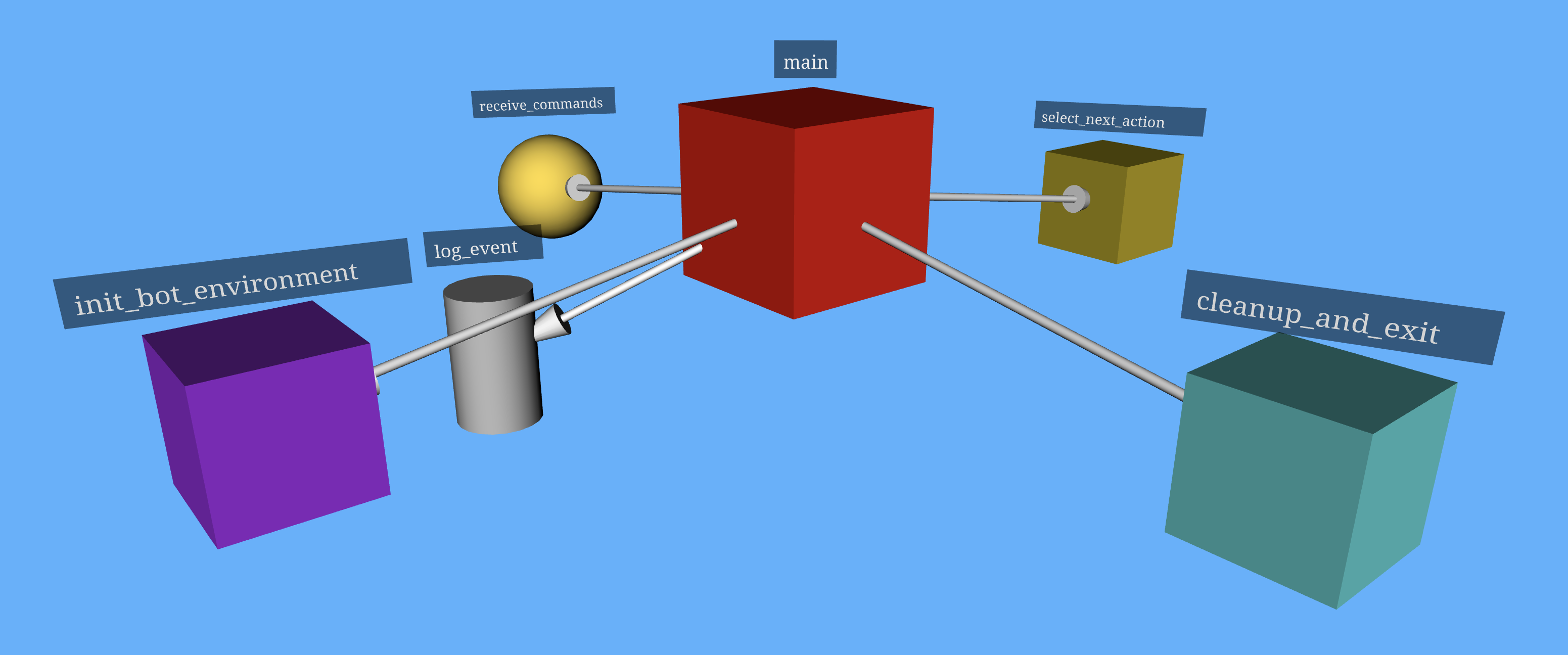}}
        \caption{Lowest-rated call graph for v11 (subjective): High guidance, GPT-4.1}
        \label{fig:lowest_sub_v11}
    \end{subfigure}
    \hfill
    \begin{subfigure}[b]{0.49\textwidth}
        \centerline{\includegraphics[width=0.50\linewidth]{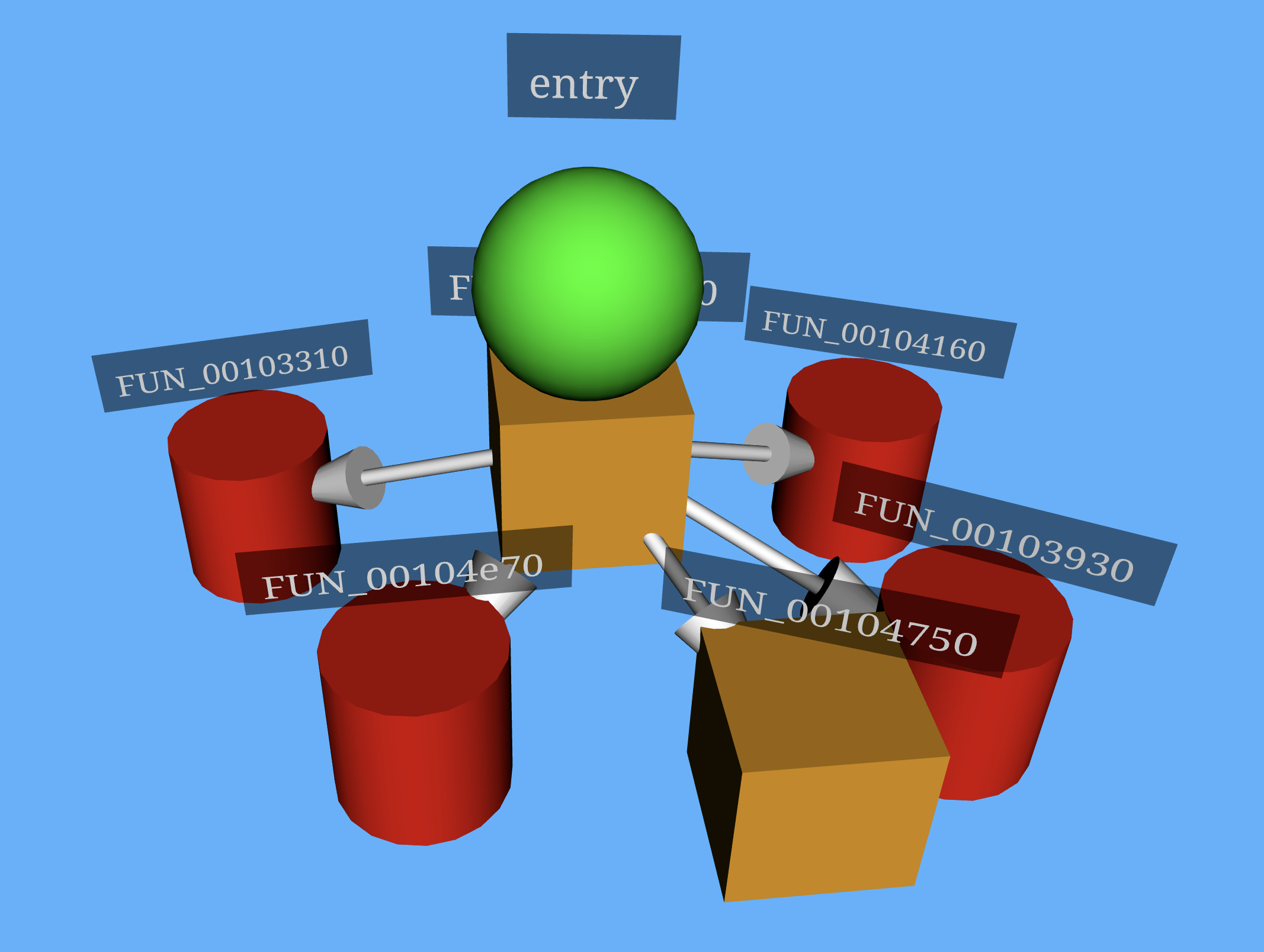}}
        \caption{Lowest-rated call graph for hexdump (subjective): High guidance, o4-mini}
        \label{fig:lowest_sub_hexdump}
    \end{subfigure}
    
    \caption{Comparison of call graphs for v11 and hexdump.}
    \label{fig:graph_comparison}
\end{figure*}

To assess variability, we first considered the objective measures and computed a composite diversity score using the average of normalized values as shown in Table~\ref{table:results_summary}. The most diverse configuration was ``v11'' + high guidance + o4-mini, while the least diverse was ``hexdump'' + low guidance + GPT-4.1. Notably, when considering subjective ratings, the same configuration produced both the highest- and lowest-rated graphs for both programs (`v11'' + high guidance + GPT-4.1 and ``hexdump'' + high guidance + o4-mini). Those outputs are shown in Figure~\ref{fig:graph_comparison} along with the basic 2D call graphs of each program. 

Subjectively, raters preferred graphs that used spatial grouping, minimized clutter, and replaced or augmented obfuscated names with inferred semantic names. These observations suggest that LLMs include some latent alignment with perceptual and task-relevant design goals, even without explicit visualization training.

\begingroup
\setlength{\tabcolsep}{2pt} 
\renewcommand{\arraystretch}{1} 
\begin{table}[tbp]
\centering
\begin{tabular}{|lll|l|l|l|l|}
\hline
\rowcolor[HTML]{C0C0C0} 
\multicolumn{1}{|l|}{\cellcolor[HTML]{C0C0C0}Program} & \multicolumn{1}{l|}{\cellcolor[HTML]{C0C0C0}\begin{tabular}[c]{@{}l@{}}Guidance \\ Level\end{tabular}} & LLM & \begin{tabular}[c]{@{}l@{}}Avg Subj \\ Rating \\ Mean 1-5\end{tabular} & \begin{tabular}[c]{@{}l@{}}Avg Subj \\ Rating \\ CV\end{tabular} & \begin{tabular}[c]{@{}l@{}}Composite \\ Obj Score \\ Mean 0-1\end{tabular} & \begin{tabular}[c]{@{}l@{}}Composite \\ Obj Score \\ CV\end{tabular} \\ \hline
\multicolumn{1}{|l|}{hexdump} & \multicolumn{1}{l|}{high} & gpt41 & 3.15 & 0.12 & 0.29 & 0.2 \\ \hline
\multicolumn{1}{|l|}{hexdump} & \multicolumn{1}{l|}{high} & o4mini & 2.67 & 0.24 & 0.16 & 0.31 \\ \hline
\multicolumn{1}{|l|}{hexdump} & \multicolumn{1}{l|}{low} & gpt41 & 2.47 & 0.2 & 0.24 & 0.3 \\ \hline
\multicolumn{1}{|l|}{hexdump} & \multicolumn{1}{l|}{low} & o4mini & 2.72 & 0.13 & 0.21 & 0.43 \\ \hline
\multicolumn{1}{|l|}{v11} & \multicolumn{1}{l|}{high} & gpt41 & 2.96 & 0.21 & 0.33 & 0.32 \\ \hline
\multicolumn{1}{|l|}{v11} & \multicolumn{1}{l|}{high} & o4mini & 3.22 & 0.15 & 0.46 & 0.52 \\ \hline
\multicolumn{1}{|l|}{v11} & \multicolumn{1}{l|}{low} & gpt41 & 2.91 & 0.13 & 0.33 & 0.21 \\ \hline
\multicolumn{1}{|l|}{v11} & \multicolumn{1}{l|}{low} & o4mini & 3.35 & 0.05 & 0.21 & 0.26 \\ \hline
\multicolumn{3}{|l|}{ALL hexdump} & 2.75 & 0.18 & 0.22 & 0.36 \\ \hline
\multicolumn{3}{|l|}{ALL v11} & 3.11 & 0.15 & 0.33 & 0.46 \\ \hline
\multicolumn{3}{|l|}{ALL High Guidance} & 3.0 & 0.18 & 0.31 & 0.54 \\ \hline
\multicolumn{3}{|l|}{ALL Low Guidance} & 2.86 & 0.16 & 0.25 & 0.34 \\ \hline
\multicolumn{3}{|l|}{ALL gpt41} & 2.87 & 0.18 & 0.3 & 0.27 \\ \hline
\multicolumn{3}{|l|}{ALL o4mini} & 2.99 & 0.17 & 0.26 & 0.66 \\ \hline
\end{tabular}
\caption{Mean and Coefficient of Variation for Subjective and Objective Ratings}
\label{table:results_summary}
\end{table}
\endgroup

\section{Conclusion and Future Work}
\label{sec:conclusion}

While our pilot evaluation offers preliminary indications that LLMs can sometimes produce 3D call graph layouts aligned with cognitive and perceptual design principles, the results also reveal substantial variability, including visualizations that were subjectively less effective or ineffective regardless of guidance level, model, or program. We therefore regard this work as an initial exploration rather than confirmation that current LLMs possess cognitive awareness in visualization. No causal link has been established between the model’s training data and the effectiveness of the generated layouts, and further investigation is needed to understand the origins of any cognitively aligned patterns that do emerge. We found additional practical limitations with the LLM interactions: programs having more than tens of functions hit context window and token rate limitations, so we haven't been able yet to try this approach on large binary programs; and even with small binaries, each call graph required approximately 10 to 300 seconds to generate, which does not support true interactivity. Finally, rating the models in a desktop model viewer may produce different results than if the participants rated models in the immersive VR environment.

Future work includes several avenues. First, integrate the MCP client into the VR interface for seamless prompting and iteration including speech-to-text and text-to-speech. Second, explore an actor-critic approach where a secondary model evaluates generated visualizations for cognitive fit. Third, determine an iterative method to support more complex binary programs. Finally, there is ample opportunity to more deeply study prompting approaches, especially to address the variation in results; rather than add fixed limitations on the LLMs, use more implicit and explicit observations of the user's cognitive and perceptual state to help focus the LLM's generation of visualizations.

Overall, this study provides an early demonstration of LLM-driven visualization in immersive reverse engineering, identifies key challenges, and outlines a path toward more reliable and cognitively aligned visualization agents.

\bibliographystyle{IEEEtran}
\bibliography{VR}

\end{document}